# A New Approach to Spreadsheet Analytics Management in Financial Markets

Brian Sentence, Xenomorph Software Ltd

bsentance@xenomorph.com

**ABSTRACT**

*Spreadsheets in financial markets are frequently used as database, calculator and reporting application combined. This paper describes an alternative approach in which spreadsheet design and database technology have been brought together in order to alleviate management and regulatory concerns over the operational risks of spreadsheet usage. In particular, the paper focuses on the rapid creation and centralised deployment of statistical analytics within a software system now in use by major investment banks, and presents a novel technique for the manipulation in spreadsheets of high volumes of intraday market data.*

## 1 INTRODUCTION

Whilst spreadsheets are used extensively in many professions, nowhere is this usage more pervasive or more critical than in the financial markets [*Croll, 2005*]. There are many factors that contribute to why this is the case. Client, regulatory and competitive pressures are forcing trading desks to analyse ever-higher volumes of market data in order to demonstrate customer value, to show adherence to market rules and to identify new trading opportunities. The complexity and breadth of the data being analysed is also increasing, due to the innovative nature of the new financial products that are now being created.

Another key factor driving the usage of spreadsheets in financial markets is the relative extremes of specialisation within the industry. End-users such as derivatives traders, product controllers, risk managers and quantitative analysts ordinarily need to have a very firm foundation in the understanding of mathematics, financial theory and market behaviour. This specialisation of end-user knowledge, combined with extremely short commercial delivery timeframes, presents a huge challenge to even the most advanced of system designers. This naturally leads to increased spreadsheet usage as the spreadsheet becomes the only platform that can meet the deadlines imposed by market and competitive pressures.

The background described above would only be of passing importance if it were not for the huge sums of money being managed out of spreadsheets on a day-to-day basis. One seemingly small error in a trader's spreadsheet could potentially cause (and has caused) very significant losses for a financial institution [*Wilmott, 2005*]. Even in the absence of any spreadsheet errors, due to a spreadsheet's lack of transparency then a financial institution is open to the operational risk that an individual trader may deliberately mis-quote or mis-represent the instrument pricing and risk levels being undertaken [*Mittemeir, 2005*]. It is therefore no surprise that regulators are now paying very direct attention to the use of spreadsheets by banks [*Buckner, 2004; PWC, 2004*]

This paper describes a new approach to this problem of spreadsheet management in financial markets. It combines the best of spreadsheet productivity with the best of database technology to provide consistent, centralised and transparent access to data for all users. Additionally, an object spreadsheet approach is described which can greatly reduce the number of spreadsheet formulas required to manipulate large amounts of array data. The approach taken puts spreadsheet design at the heart of the data management process for a financial institution rather than as an "ad-hoc" or tactical add-on solution.

## 2 SOLUTION DESIGN GOALS

Whilst many of the spreadsheet issues described above are as much procedural as technical, it is technically possible to address many of the negative sides of their usage whilst also leaving many of their positives aspects in place. This paper describes a very recent spreadsheet-related enhancement to a data management system that it is currently in use at some of the major global investment banks. Traders and risk managers use the system, known as TimeScape, to perform statistical analysis on historical market data and to perform derivatives valuation in both pre- and post-trade decision support. What follows below are some of the key design considerations of this centralised spreadsheet environment known to the author as a "Formula Grid" and illustrated in Figure (1).

### 2.1 Spreadsheet Interface

One major design goal was to ensure that users could still benefit from the productivity of using a spreadsheet interface to define calculations and analytics. This is particularly vital in financial markets where specialist business knowledge combined with commercial timeframes often preclude the transfer of this knowledge to system designers. Hence Formula Grid calculations can be edited by end-users alongside of their usage of spreadsheets and other applications. This is shown in Figure (1) as the Formula Grid Editor.

A further step was also taken to make the Formula Grid spreadsheet calculation an in-line, core part of the process of market data management and analysis, rather than an ad-hoc tool into which data is imported, analysed and exported out into the business process. This was implemented by means of mapping and hiding spreadsheet calculations behind analytical functions and data fields, more of which will be seen later in this paper.

### 2.2 Data Centralisation

Here the approach taken was to move the data out of the spreadsheet and to locate the data within a centralised database as shown in Figure (1). There are a few key considerations when doing this. The first is to make sure that the database can support the typical data types used in spreadsheets such as arrays, lists and matrices. These datatypes are not typically found out of the box with any traditional database management systems. The second is to make it easy for the user to get data from existing spreadsheets into the new centralised spreadsheet environment. Once again, databases tend to be very technical in nature so this naturally alienates the community of end-users that we wish to assist.

Given that these two aspects are in place it is then possible to reduce the amount of data actually stored "within" the spreadsheet environment (e.g. "cell = data value" ) and allow more of this data available outside of the spreadsheet for other non-spreadsheet users. Once the data is contained within a centralised database it is then easy to make use of the usual things that databases do well such as user access permissioning, backing up and restoring data, providing transparent programming access to data and so on.

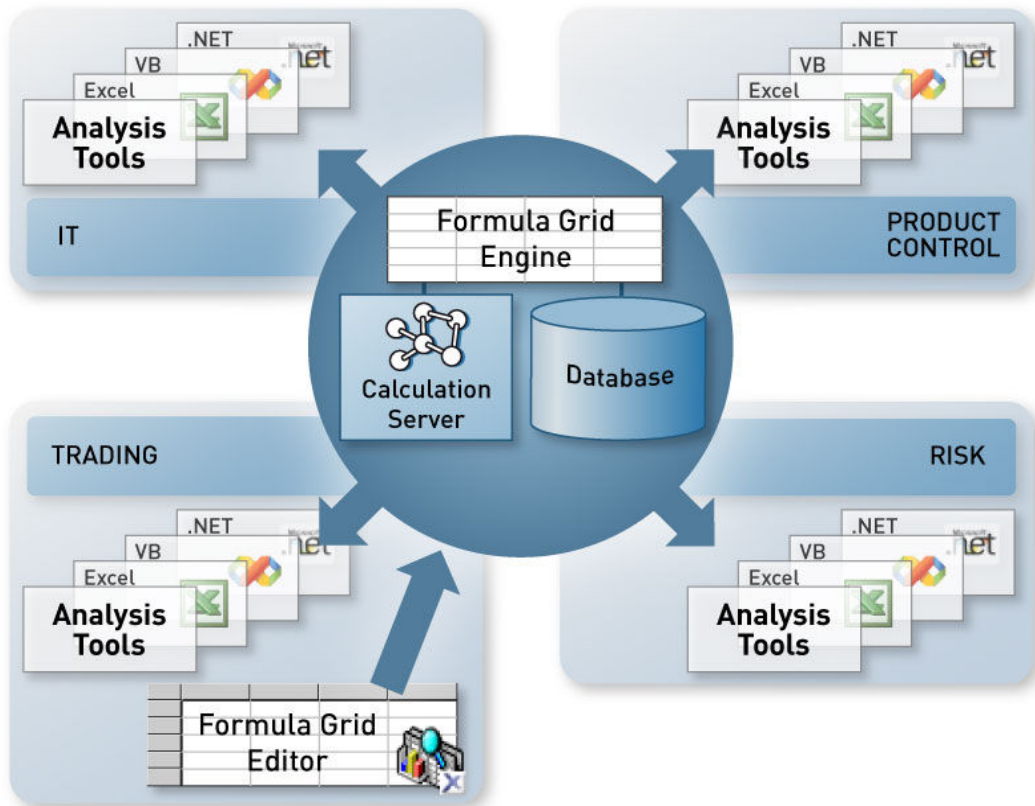

**Figure (1) – Server-Side Spreadsheet Calculation with Client-Side Editing**

*2.3 Calculation Centralisation*

Whilst end-users like to use spreadsheet interfaces to define data and calculations, this in itself does not preclude the possibility of the calculation being defined by the user being run in a centralised, server-side manner. This is approach taken here, where spreadsheet operations defined by the user are stored centrally in the database in Figure (1) and also run centrally in the calculation server of Figure (1). This means that these spreadsheet calculations are available not just to end users of spreadsheets but to all users throughout an organisation, as again illustrated by Figure (1). Calculation centralisation also leaves the architecture open to further improvement as software and hardware infrastructure improve, all without the end user needing to be aware of any material changes other than improved performance.

*2.4 Data Objects in Cells*

In order to cope with large amounts of array data found in financial markets, an additional behaviour was introduced. This allowed array and other more complex data types to be contained within a single spreadsheet cell. For example, a time series of bond or equity prices could be contained within a single spreadsheet cell (e.g. cell "A1 = Closing Price Series" or cell "B1 = Historic FX Series"). This then allows vector arithmetic, such as converting the historic prices of an equity from one currency into another, to be defined through simple cell operations such as cell C1 = A1 * B1. Such an example operation would take all of the historic equity prices contained in cell A1, all the historic FX rates stored in cell B1 and multiple them together throughout all time to produce an array result of the correct currency in cell C1.

## 3 SOLUTION EXAMPLE

What follows is an example of calculating a Volume Weighted Average Price (VWAP) measure which is often used by portfolio traders to demonstrate to clients how well a client's order to sell or buy stock has been placed in the market against average price levels observed. The example shown is greatly simplified in order to best illustrate the principles of how a centralised spreadsheet calculation is defined and executed as a Formula Grid within the data management environment utilised. VWAP practitioners should note that much more complex, flexible and parameterised versions of VWAP calculation can also be implemented in the Formula Grid, but these are outside of the scope of this introductory paper.

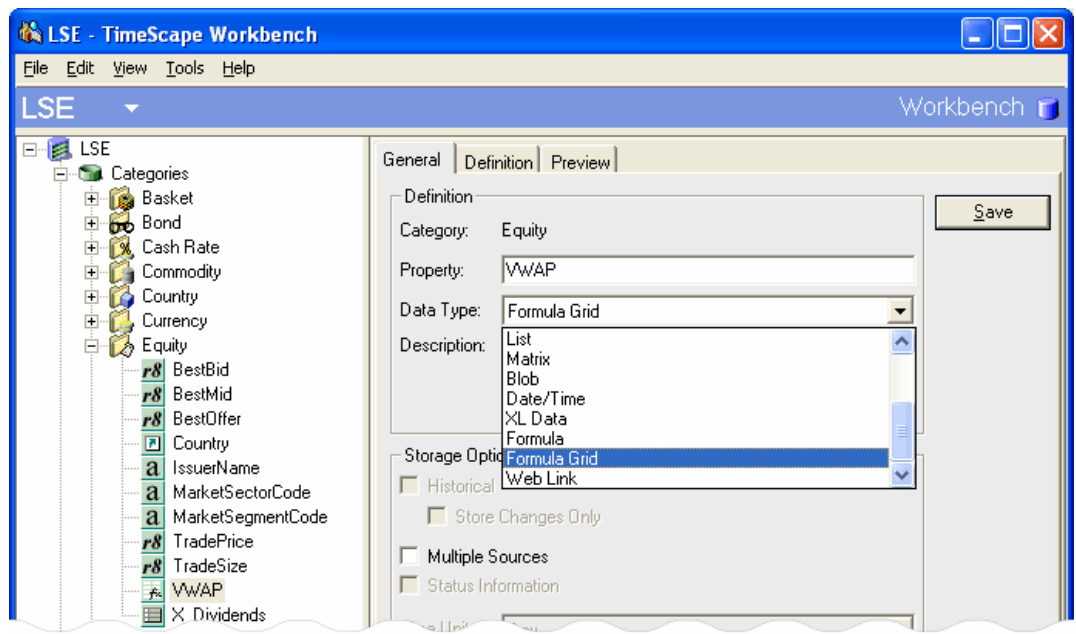

**Figure (2) Creating a Formula Grid Data Attribute for Equity Instruments**

Figure (2) above shows the schema (standard data attributes) for all "Equity" instruments contained in a database of market and static data known as LSE. Already set up for equity instruments are data attributes such as "TradePrice" and "TradeSize", both of which are numeric time series stored externally to the Formula Grid calculation we are about to design. In particular, Figure (2) illustrates how a new data attribute is being created called "VWAP" and its data type is being assigned to type Formula Grid from the dropdown shown. After the attribute has been created, a tab appears containing a spreadsheet environment upon which calculations can be defined. In the example shown in Figure (3) below, an equity attribute of "TradePrice" has been placed in cell A1.

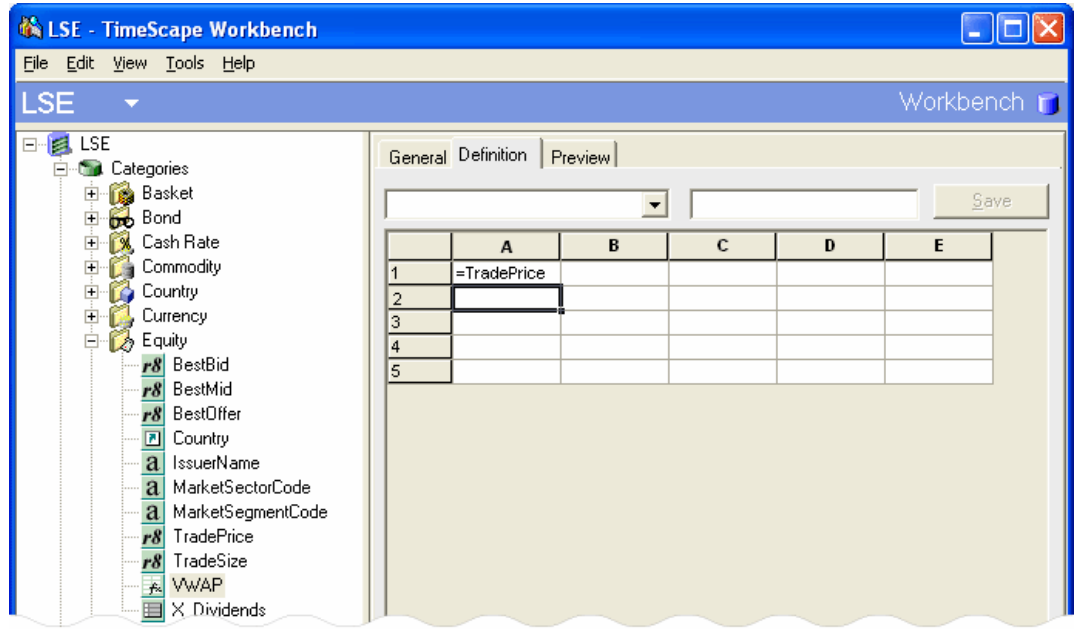

**Figure (3) Entering the Spreadsheet Calculation in a Formula Grid**

Given that this Formula Grid calculation is centralised and is going to apply to all equity instruments contained in the database, then it is useful to preview the output for an example equity. Boots PLC has been chosen as the example equity and in Figure (4) below it shows how 25,000 prices from the "TradePrice" attribute of Boots PLC are being accessed in one Formula Grid spreadsheet cell.

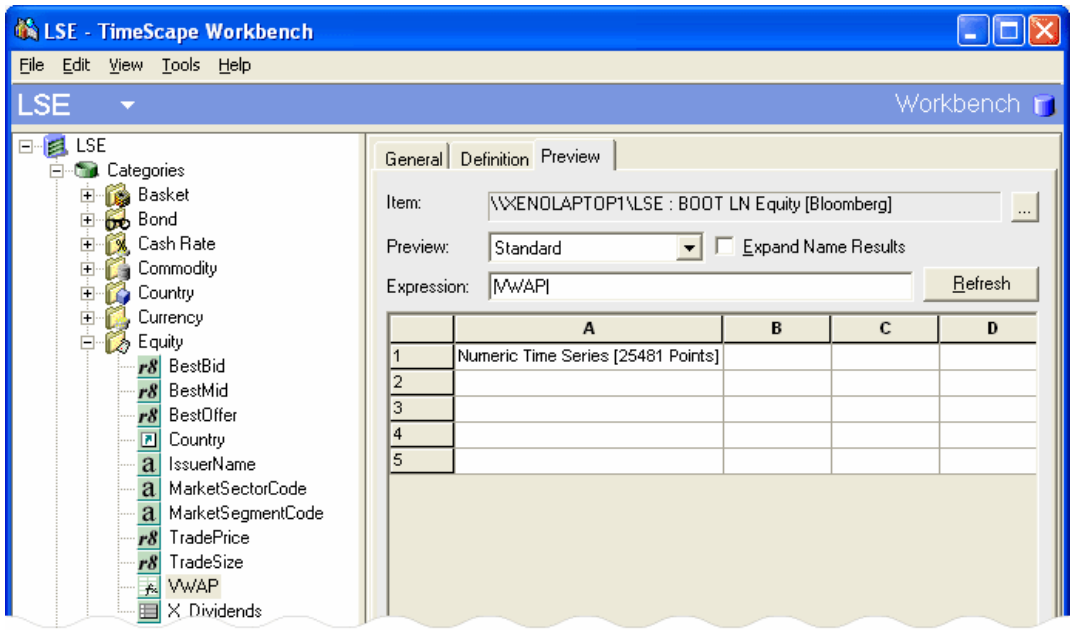

**Figure (4) 3D Preview of Array Data Containing in Cells**

Whilst the preview in Figure (4) may be a useful way of representing data as part of manipulating it, then Figure (5) shows how the data contained in cell A1 can be unfolded to present a more traditional and indeed "human-readable" representation of "TradePrice" as a two column array of times and recorded prices for Boots PLC.

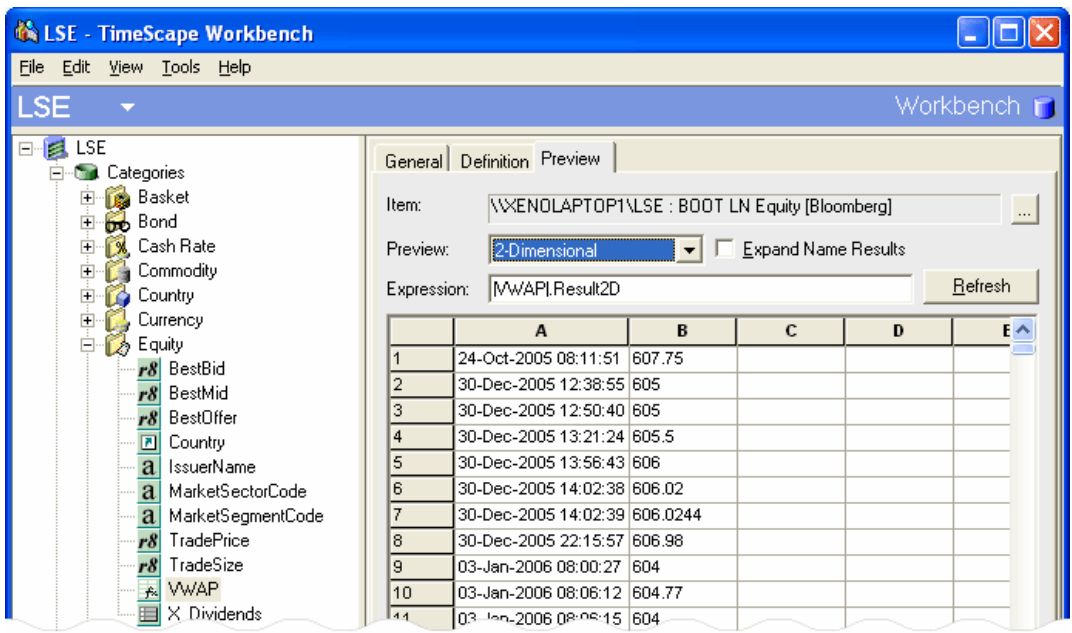

**Figure (5) Unfolded 2D Preview of Array Data Containing in Cells**

Now that we can see how arrays can be accessed within single spreadsheet cells in the Formula Grid, it is now possible to define a simple spreadsheet formula for the VWAP calculation as shown in Figure (6) below.

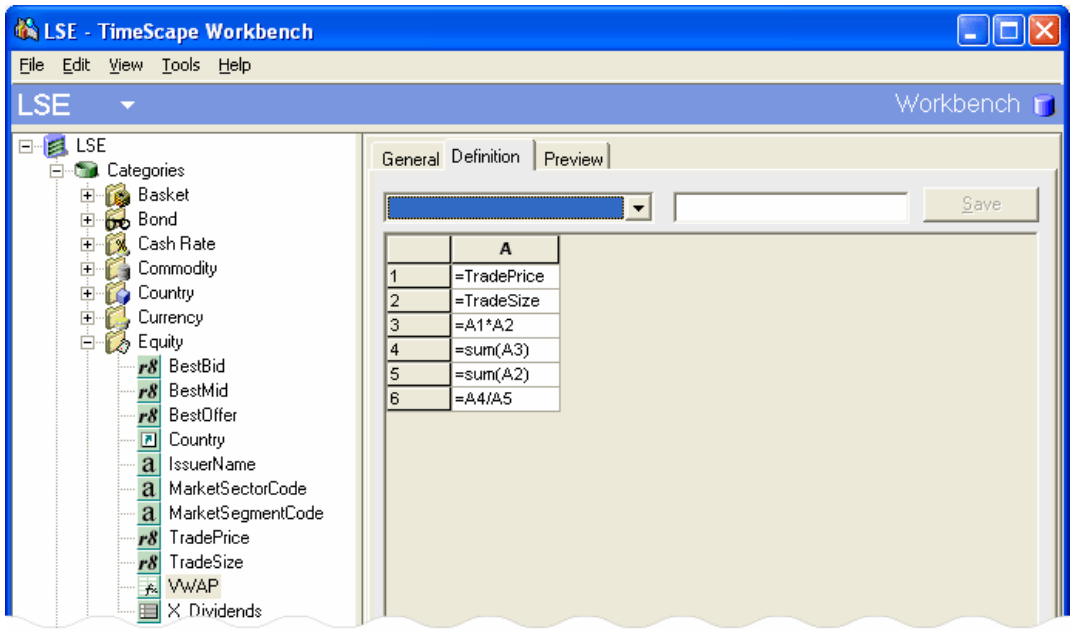

**Figure (6) VWAP Calculation Showing Vector Multiplication**

Figure (6) shows how the "TradePrice" has been referenced in cell A1 and the "TradeSize" in cell A2. In cell A3 we multiple the price and the volume cells together to produce a vector result. In cells A4 we sum all of the historic array values in A3 to produce a scalar result, and similarly we sum the total volume of all transactions from cell A2 in cell A5. Finally, we divide the product of price and volume by the total volume to give us the VWAP result in cell A6.

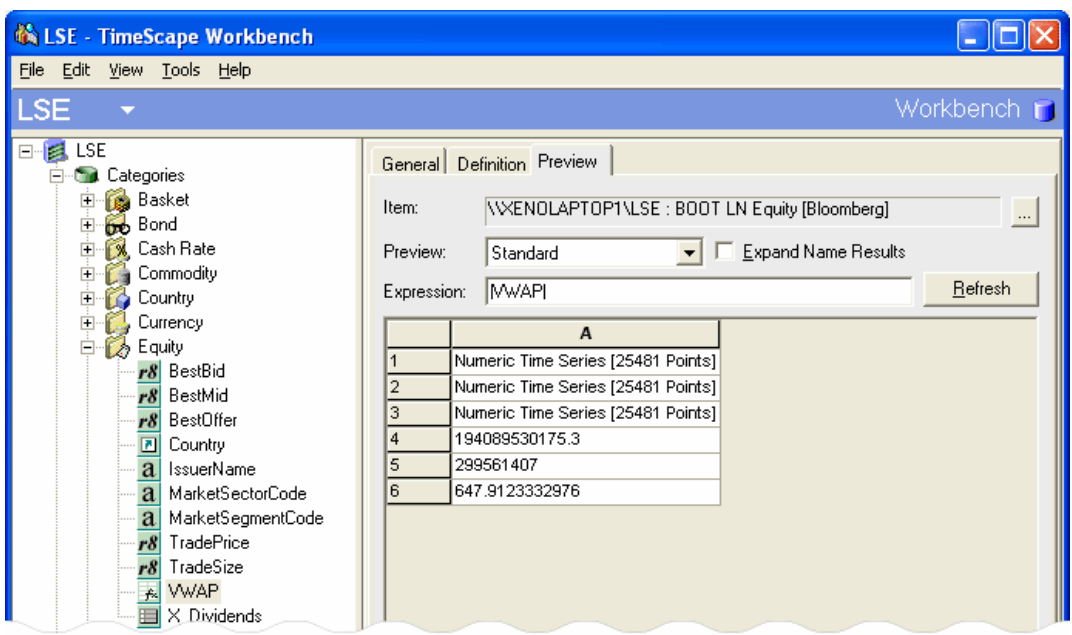

**Figure (7) 3D Preview VWAP Calculation Results**

Figure (7) above shows the results preview (in array form) of executing the spreadsheet calculation defined in Figure (6) for Boots PLC. The first three cells contain vector results and the last three contain scalars, as expected from the formula definition.

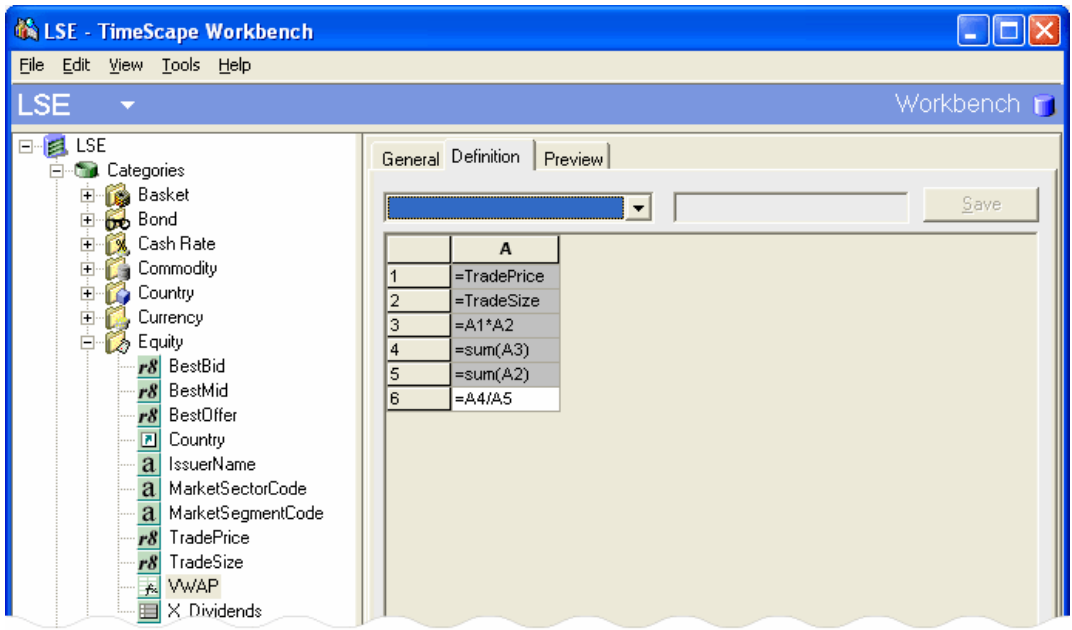

**Figure (8) Hiding Intermediate VWAP Calculations**

Figure (8) shows that the intermediate calculations in the first five cells, A1 to A5, can be hidden from the end user and this is shown by the cells being greyed to mark their "hidden" status. One passing point of note is that when multiplying time vectors to produce the result in cell A3, it may prove necessary to align data through time which can be handled by an extensive set of data rules built into the Formula Grid Engine.

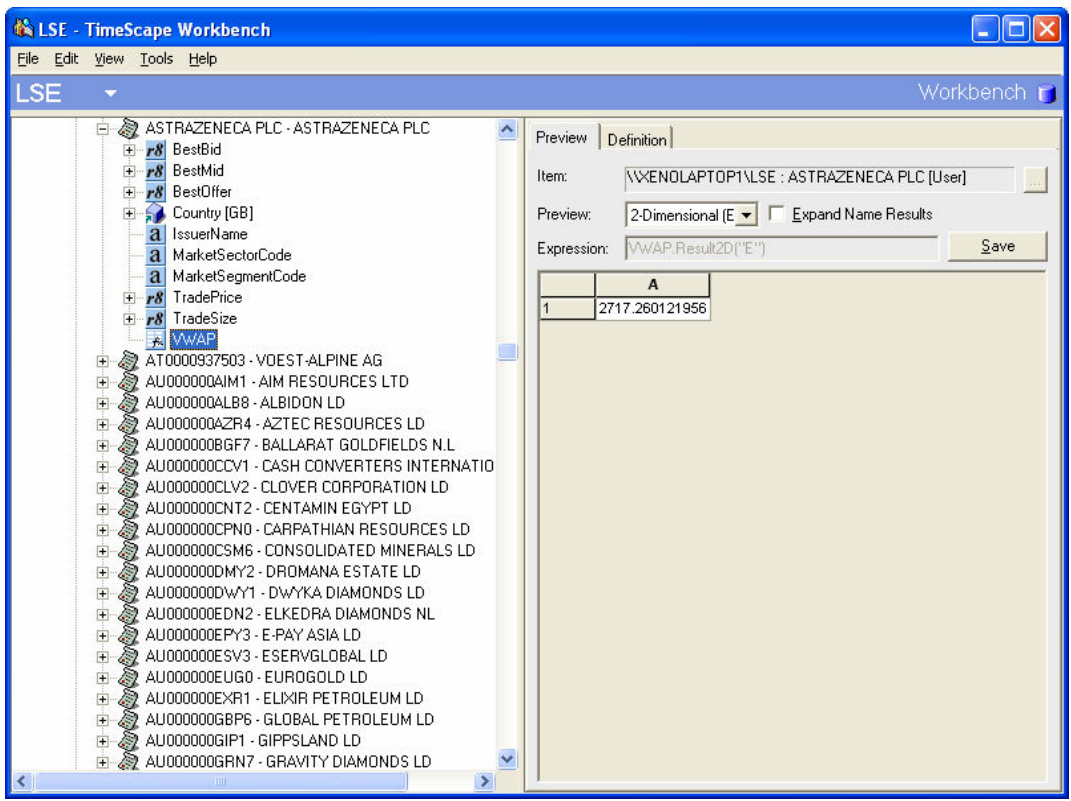

**Figure (9) Viewing VWAP Results For Any Equity**

In Figure (9) above we move away from the database schema and the preview of a single instrument with Boots PLC shown in previous screen shots, in order to browse and view an entire data universe of equity instruments contained within an equity database. On browsing to find the equity AstraZenica PLC we then select the data attribute called "VWAP", which causes the Formula Grid spreadsheet calculation defined in Figure (6) to execute in the context of the particular equity being viewed. Hence AstraZeneca PLC's "TradePrice" and "TradeSize" series have been loaded in background and the VWAP calculated as a value of 2717.

All of this has occurred without the end user needing to be aware of the complexity of the calculation or the way in which it was defined, and is available to all users of the system whether browsing equity data as above or loading data through programming interfaces.

## 4 CONCLUSION

This paper has presented an alternative method to the usage of spreadsheets as database, calculator and reporting application combined. The approach described is already in use within a commercial data management software system, and combines spreadsheet design with database technology to achieve centralised and transparent deployment of financial analytics. In particular it has illustrated a novel technique for the manipulation of large amounts of array data, which is becoming more important in financial markets as practitioners become increasingly interested in the analysis high-frequency intraday market prices and quotes.